\documentclass[review]{elsarticle}

\usepackage{xr}
\usepackage{hyperref}

\usepackage{graphicx}
\usepackage{multirow}
\usepackage{placeins}
\usepackage{array}
\usepackage{amsmath}
\usepackage{lineno}
\usepackage[margin=1in]{geometry}
\usepackage{color}
\usepackage{soul}

\usepackage{ragged2e}

\journal{Energy Policy}

\begin{document}

% \begin{frontmatter}

% %% Group authors per affiliation:
% \author{Elsevier\fnref{myfootnote}}
% \address{Radarweg 29, Amsterdam}
% \fntext[myfootnote]{Since 1880.}

% \end{frontmatter}

% \linenumbers
% \def\linenumberfont{\normalfont\scriptsize\sffamily}

\begin{frontmatter}

\title{Exploratory analysis of high-resolution power interruption data reveals spatial and temporal heterogeneity in electric grid reliability}

% \author{Laurel N. Dunn, Michael D. Sohn, Kristina Hamachi Lacommare \& Joseph H. Eto}

% \maketitle

%% or include affiliations in footnotes:
\author[berkeley,lbnl]{Laurel N. Dunn\corref{correspondingauthor}}
\cortext[correspondingauthor]{Corresponding author}
\ead{lndunn@berkeley.edu}

\author[lbnl]{Michael D. Sohn}
% \ead{mdsohn@lbl.gov}

\author[lbnl]{Kristina Hamachi LaCommare}
% \ead{kshamachi@lbl.gov}

\author[lbnl]{Joseph H. Eto}
% \ead{jheto@lbl.gov}

\address[berkeley]{Department of Civil and Environmental Engineering, University of California Berkeley}
\address[lbnl]{Energy Technologies Area, Lawrence Berkeley National Laboratory}

\begin{abstract}

Modern grid monitoring equipment enables utilities to collect detailed records of power interruptions.
These data are aggregated to compute publicly reported metrics describing high-level characteristics of grid performance.
The current work explores the depth of insights that can be gained from public data, and the implications of losing visibility into heterogeneity in grid performance through aggregation.
We present an exploratory analysis examining three years of high-resolution power interruption data collected by archiving information posted in real-time on the public-facing website of a utility in the Western United States.
We report on the size, frequency and duration of individual power interruptions, and on spatio-temporal variability in aggregate reliability metrics.
Our results show that metrics of grid performance can vary spatially and temporally by orders of magnitude, revealing heterogeneity that is not evidenced in publicly reported metrics.
We show that limited access to granular information presents a substantive barrier to conducting detailed policy analysis, and discuss how more widespread data access could help to answer questions that remain unanswered in the literature to date.
Given open questions about whether grid performance is adequate to support societal needs, we recommend establishing pathways to make high-resolution power interruption data available to support policy research.

\end{abstract}

\begin{keyword}
Electric power systems\sep Electric power interruptions\sep Grid reliability metrics \sep Big data \sep Data access \sep Exploratory data analysis
\end{keyword}

\end{frontmatter}

\section{Introduction}
\label{sec:intro}

With modern monitoring and control equipment, electric power systems deliver high levels of service reliability to grid customers.
Monitoring systems provide detailed data about the power interruptions that do occur and the customers who are affected.
Power interruption data can support empirical research exploring a wide range of policy questions relevant to how we assess, regulate, and invest in grid performance.

These granular data are proprietary, but are aggregated to compute publicly reported reliability metrics.
Information regarding the characteristics of individual power interruptions is lost through aggregation.
However, the implications of losing this information are not well understood.
The current work explores whether (or not) the loss of granular information is an impediment to using the available data to support policy analysis and research.

Several works in the literature to date use high-resolution power interruption data to explore a wide range of research questions.
Examples include weighing different tree trimming policies \cite{guikema2006}, and using predictive models to inform preventative maintenance strategies \cite{rudin2010}.
Vulnerabilities to hurricanes \cite{nateghi2014a, ji2016} and other weather conditions \cite{kankanala2014} are also explored.
However, these studies rely on proprietary data that are not readily available.
As such the geographic scope is generally limited to a single service territory.
Operational, topological, and climatic differences between service territories raise questions about how generalizable results are to other geographic regions.

Research efforts that are regional or national in scope require datasets that are more geographically comprehensive.
In light of the barriers to securing proprietary data from multiple utilities, publicly reported information are better suited to support these efforts.
Several studies in the literature aim to capitalize on these datasets, for example to report on trends in grid performance \cite{hines2009, larsen2015, amin2008}, to quantify the implications of power interruptions \cite{lacommare2004} and to explore the reliability benefits of different undergrounding policies \cite{larsen2016}.
Further efforts have also been made to develop explanatory models for correlating reliability metrics with local weather patterns \cite{larsen2015, caswell2011}.

Several of these studies leave questions unanswered, citing low-resolution in the data they use as both a limitation of the analysis they performed and a barrier to future work \cite{hines2009, larsen2015, larsen2017}.
This raises questions about whether there are policy implications associated with leaving these research questions unanswered, whether more granular reliability data could advance policy research, and if so whether collection of more granular data is warranted.

Multiple studies examining national grid performance report that the frequency of power interruptions is increasing \cite{hines2009, larsen2015, amin2008}.
This apparent decline in grid performance coincides with an increase in grid infrastructure spending per unit of electricity sold. 
Between 1990 and 2015, the ratio between infrastructure expenditures (in transmission and distribution systems combined) and electricity sales doubled \cite{eia2012, eia2001}.

Reports that reliability is decreasing suggest that there is some uncertainty about whether or not investments in grid performance have improved the reliability of service to grid customers.
Further research is needed to understand what is causing the observed trends, to evaluate whether (or not) there is cause for concern, and to determine what policy interventions (if any) are needed to mitigate a decline in grid performance.

The reported change in grid reliability may not be a cause for concern if grid performance is improving in ways that these aggregate metrics do not capture.
For example, investments in grid performance can target measures that benefit those subsets of customers that are most vulnerable to power interruptions.
Though such measures may improve grid performance by some measures, these gains are not necessarily evidenced by metrics that describe the characteristics of power interruptions and not their implications.

An active body of research is exploring quantitative methods for evaluating these reliability gains.
For example, damage functions estimate costs incurred due to power interruptions by individual grid customers \cite{sullivan2015} and nationwide \cite{lacommare2004}.
Downtime of critical loads such as hospitals and water treatment facilities have also been considered \cite{nagarajan2016}.
Applying these methods to assess the implications of past power interruptions is a critical step towards evaluating whether (or not) the apparent decline in grid performance is cause for concern.

Studies consistently report that the implications of power interruptions are related to when and where the power interruption occurs.
For example, Carlsson and Martinsson \cite{carlsson2008} and Chang et al \cite{chang2007} examine the economic damages due to specific power interruptions in Canada and Sweden.
Both studies report that weather exacerbated damages, as the interruptions occurred during severe winter storms and affected customers who relied on electricity to heat their homes.
Sullivan et al \cite{sullivan2015} also show that the damages a given customer incurs vary depending on time of day and time of year.
These studies show that accurately characterizing damages requires insight into temporal heterogeneity in grid performance.

Furthermore, the characteristics of power interruptions may change depending on the loads and services that are affected.
In an analysis of proprietary power interruption data, Maliszewski and Perrings \cite{maliszewski2011} find that residential customers located near hospitals experience characteristically shorter power interruptions than similar customers located elsewhere.
The implication is that aggregate metrics likely do not represent the typical characteristics of power interruptions affecting critical loads. 
This finding shows that to evaluate grid performance requires insight into geographic heterogeneity in power interruption characteristics.

In the current work, we explore questions related to if and how granular power interruption data could advance data-driven research informing how we understand grid performance.
We do so by presenting an exploratory analysis of spatially and temporally resolved power interruption data.
We collected these data by archiving real-time information posted on public-facing utility websites about ongoing power interruptions.
We mine the data to explore heterogeneity in the characteristics of power interruptions, and report on the patterns we observe.
We compare our findings with the information contents of publicly available reliability data, and with findings from the literature reporting on these public data.
We discuss whether (or not) past studies have arrived at similar insights from analyzing public data, and whether replicating past studies with more granular data could change results reported in the literature to date.
Results showing the two datasets yield similar insights indicate that there is little marginal benefit to collecting additional data.
However, in many cases our data reveal new insights that are not evidenced in public datasets.
This finding suggests that access to more granular data may be able to resolve questions that remain unanswered in the literature to date (such as if and why grid performance is decreasing), and support new research related to grid performance.

This paper is structured as follows: Section \ref{sec:background} describes publicly available information about grid reliability and surveys the literature that draws on these datasets. Section \ref{sec:methods} describes the dataset we collected to support the current work and methods for calculating conventional reliability metrics from these data. Section \ref{sec:results} summarizes our results and discusses the implications of these results in the context of the literature to date. Finally, Section \ref{sec:conclusion} summarizes our findings and discusses opportunities to capitalize on granular datasets to inform policy interventions for improving grid performance.

\section{Publicly available power interruption data}
\label{sec:background}
Three main sources of power interruption data are readily available to the public.
These include: (1) aggregate metrics of grid reliability, (2) incident reports describing specific reliability events that are of particular interest to regulators, and (3) real-time information about ongoing power interruptions posted on public-facing utility websites.

The first two data sources are published by regulatory agencies who periodically collect information from electric utilities to track grid performance.
At the national level, the Energy Information Administration (EIA) collects annual reliability metrics \cite{eia861}, and the Department of Energy's (DOE) Office of Electricity (OE) collects information on specific reliability incidents \cite{oeform417}.
Local public utilities commissions often collect data similar to those reported to EIA and DOE, though local reporting requirements may vary.
Submission of emergency incident reports to DOE is mandatory, while reporting annual reliability metrics to EIA is voluntary.
Reporting requirements may also include sensitive information that are ultimately withheld from publicly available versions of the data.

The EIA collects data on reliability, along with a wide range of information about utility sales and operations via EIA form 861 \cite{eia861}.\footnote{Available for download here: https://www.eia.gov/electricity/data/eia861/ (Accessed November, 2018)}
Utilities submit this form annually.
Three reliability metrics are reported: System Average Interruption Frequency Index (SAIFI), System Average Interruption Duration Index (SAIDI) and Customer Average Interruption Duration Index (CAIDI).
SAIFI describes the average number of power interruptions customers experience per calendar year.
SAIDI describes the average amount of time (in minutes) each customer spends without electricity service per calendar year.
CAIDI describes the average duration (in minutes) of individual service interruptions.
Standard definitions of these metrics and methods for computing them are detailed in IEEE Standard 1366 \cite{ieee2012}.

Though EIA only recently began collecting these data at a national scale, many local public utilities commissions have collected similar data for a longer time.
Several works in the literature study data collected by local reporting agencies.
For example, Larsen et al \cite{larsen2015} report on longitudinal SAIDI and SAIFI data to examine long-term trends in grid reliability. 
LaCommare and Eto \cite{lacommare2006} link similar data to damage functions reported in Sullivan et al \cite{sullivan2015} to estimate the scale of damages grid customers incur due to power interruptions nationally each year.

The DOE Office of Electricity (OE) collects more detailed information describing characteristics of specific power interruptions via DOE Form OE-417.\footnote{Data are available for download here: http://www.oe.netl.doe.gov/oe417.aspx (Accessed November, 2018)}.
Power interruptions reported to OE are of particular interest due to their size (e.g., more than 50,000 customers affected) or underlying cause (e.g., cyber or physical attack), and are referred to as ``emergency incidents''.

When an emergency incident occurs, the affected utility (or balancing authority) files an Electric Emergency Incident and Disturbance Report by (via DOE Form OE-417) within 1 or 6 hours of the event, depending on the nature of the incident.
Though details about the events leading up to each incident are withheld, much of the reported information is made available to the public including:
\begin{itemize}
\item Affected geographic region
\item Start date and time
\item End date and time 
\item Peak megawatts of demand loss
\item Number of customers affected
\item Alert criteria (i.e., what caused the power interruption to classified as an emergency incident)
\item Event type/cause
\end{itemize}
We refer readers to read the form itself \cite{oeform417} for definitions of these data fields and for a complete list of the criteria that designate an emergency incident.

Finally, many electric utilities post power interruption data on their public-facing websites.
These data provide an instantaneous account of the power interruptions currently happening in their service territory.
To our knowledge there is no regulatory requirement to provide these data.
Though no explanation is given, one viable reason for publicizing the data is to provide real-time updates to grid customers who are experiencing power interruptions.
Since there is no regulatory agency overseeing the publication of these data, details such as how frequently information are updated, the minimum size (if any) of the power interruptions posted, and geographic granularity of the information provided are at the discretion of the utility maintaining the website.
Websites often state that data are updated every 15 minutes, suggesting that the process is automated.

Utilities that provide the most geographic granularity report power interruptions as point locations described by latitude and longitude coordinates; the data do not indicate the physical significance of these point locations.
Most utilities, however, report the number of affected customers aggregated by ZIP code, county, or municipality.

The current work reports on data we collected by archiving these real-time information to compile a historical record of the power interruptions that occurred.
In Section \ref{sec:data} we discuss the contents of the data and how they were collected.
While the bulk of our analysis focuses on this high-resolution data, we also compare information contents of these data with information reported in the EIA and DOE datasets.
Through these comparisons, we aim to better understand what information (if any) is lost in computing the aggregate metrics that are publicly reported.
The insights we gain allow us to comment on the analysis capabilities that granular data could afford.
\section{Methods}
\label{sec:methods}

\subsection{Data collection}
\label{sec:data}
To collect real-time information posted on utility websites, we developed an automated software tool that visits upwards of 30 utility websites on average every 9 minutes and records the current time, the locations of the power interruptions posted, and the number of customers affected at each location.
These data are archived in a PostgreSQL database.
These 30 utilities collectively serve about 40\% of electricity customers in the United States.
We have collected upwards of three years worth of data to date.

Since the current work explores the analysis capabilities of granular data, we focus on a utility that provides high geographic resolution and reports power interruptions by point location.
Focusing on a single utility further reduces heterogeneity introduced by operational or climatic differences between utilities.

Without access to utility records, we cannot confirm with certainty that the information we collect provides a precise or complete record of the power interruptions that occurred, though we do show that our observations align well with publicly reported information.
In light of these considerations, we want to make it explicit that our aim is not to generalize our results to other utilities or to precisely summarize the interruptions that occurred, but rather to explore what insights granular data can provide that public datasets do not.

We choose a utility in the Western United States that serves nearly 5 million customers across 900 ZIP codes.
Of these customers, about 93\% are residential, 6\% are commercial, and the remaining 1\% are split between industrial and agricultural customers.
We report on data collected over three years between June 1, 2014 and May 31, 2017.
The data include 16 million spatially and temporally distinct observations collected at 120 thousand distinct points in time.

\subsection{Definition of a power interruption}
\label{sec:definition}
IEEE Standard 1366 \cite{ieee2012} defines a power interruption as the complete loss of electricity service to one or more grid customers.
The duration of a power interruption is the time elapsed between when customers lose service and when service is restored.

We form an equivalent definition using the information available to us.
We define a power interruption as a series of consecutive observations (or website queries) reporting at least one grid customer without service at a specific point location.
By this definition, the number of customers affected by a power interruption can change over subsequent observations, but the location is static.

\begin{figure}[ht]
\begin{center}
\includegraphics[width=0.8\textwidth]{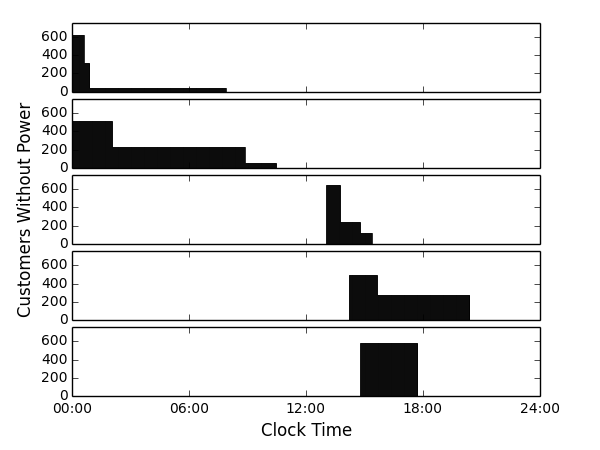}
\end{center}
\caption{Time series of customers reported to be without power at each of five point locations over the course of one day. Locations are all located in the same service territory, and were selected for having large-scale power interruptions that all occurred on the same day.}
\label{fig:event_timeseries}
\end{figure}

Figure \ref{fig:event_timeseries} shows time series data illustrating what a power interruption looks like in the data.
We show five power interruptions reported on the same day in the service territory we examine for the remainder of this work.
Each panel shows the number of customers (y-axis) without service at a specific point location over time (x-axis).

All five power interruptions affect a large number of customers compared with most of the interruptions reported in the data (as we show in Section \ref{sec:results:size}), and three of them begin at roughly the same time.
Although these observations may lead us to wonder whether there is a causal relationship between them, the data do not explicitly provide information to indicate whether or not this is the case.
In the current analysis, we designate these as five distinct power interruptions; we do not attempt to correlate them.

We define the beginning of a power interruption as the first observation reporting that customers are without service at a particular point location.
We designate the end of a power interruption as the first observation where no customers remain without service at that location.
The duration is simply the time elapsed between these two observations.

Intrinsic to this definition is the assumption that customers remain continuously without service between one observation and the next, except when a change in the data indicates that a partial or complete restoration has occurred.
Because the data are recorded in discrete sampling intervals, we cannot exclude the possibility that changes occur during the time between consecutive observations, as such changes would not necessarily be evidenced in the data.

By examining how frequently we do observe changes in the data, we can explore how frequently unobserved changes likely occur.
Across our entire dataset, the number of customers affected by a particular power interruption changes on average only once every 170 minutes.
However, we do observe some change in the overall dataset every time a new sample is recorded (every 9 minutes), for example to indicate when a new power interruption has begun.
Because the number of customers at any particular site changes much less frequently than our sampling frequency, the data suggest that unobserved changes in the number of customers affected are rare.

\subsection{Calculation of SAIDI, SAIFI and CAIDI}
\label{sec:calculations}
From the data described in Section \ref{sec:data}, we can roughly match the information contents of public reliability datasets described in Section \ref{sec:background}.
We can use these data to compute aggregate reliability metrics including SAIDI, SAIFI and CAIDI \cite{ieee2012}.

To compute SAIDI, we calculate the total minutes individual customers spend without service during the observed power interruptions and normalize by the number of customers the utility serves, as given by Equation \ref{eq:saidi_data}.

\begin{equation}
\label{eq:saidi_data}
SAIDI = \frac{1}{C_{tot}} \sum_{p \in \left\{P\right\}} \left[\sum_{i=0}^{N_p} C_{p,i} \times (t_{p,i+1}-t_{p,i})\right]
\end{equation}

\noindent
Here $C_{tot}$ describes the total number of customers the utility serves.
The set of all power interruptions ($p$) that begin during a specified time interval is given by ${P}$. 
These time intervals can be continuous (e.g., June 1, 2015) or not (e.g., between 09:00 and 10:00 AM on Weekends).
The number of observations recorded between the start and the end of power interruption $p$ is given by $N_p$.
Finally, $t_{p,i}$ and $C_{p,i}$ describe the time stamp and number of customers recorded $i$ observations after the beginning of power interruption $p$.

Graphically, the summation term for a particular interruption $p$ can also be expressed by the area of the shaded regions in each panel of Figure \ref{fig:event_timeseries}.

To compute SAIFI, we calculate the total number of customers that were affected during the observed power interruptions and normalize by $C_{tot}$, as given by Equation \ref{eq:saifi_data}.

\begin{equation}
\label{eq:saifi_data}
SAIFI = \frac{1}{C_{tot}} \sum_{p \in \left\{P\right\}} \left[C_{p,0} + \sum_{i=1}^{N_p} \Delta C\right]
\end{equation}

\noindent
Here $C_{p,0}$ describes the number of customers affected at the beginning of power interruption $p$, and $\Delta C$ describes the magnitude of all \emph{positive} changes in the number of customers affected by each power interruption, given by Equation \ref{eq:dcustomers}.
\begin{equation}
\label{eq:dcustomers}
\Delta C = \begin{cases} C_{p,i}-C_{p,i-1}, & \text{if } C_{p,i} > C_{p,i-1} \\ 0, & \text{otherwise} \end{cases}
\end{equation}

Finally, CAIDI is calculated by taking the ratio between SAIDI and SAIFI.

\externaldocument{figs}
\section{Results and Discussion}
\label{sec:results}

\subsection{Variability in interruption duration}
\label{sec:results:duration}
\subsubsection{Results}
\begin{figure}[ht]
\begin{center}
\includegraphics[width=0.8\textwidth]{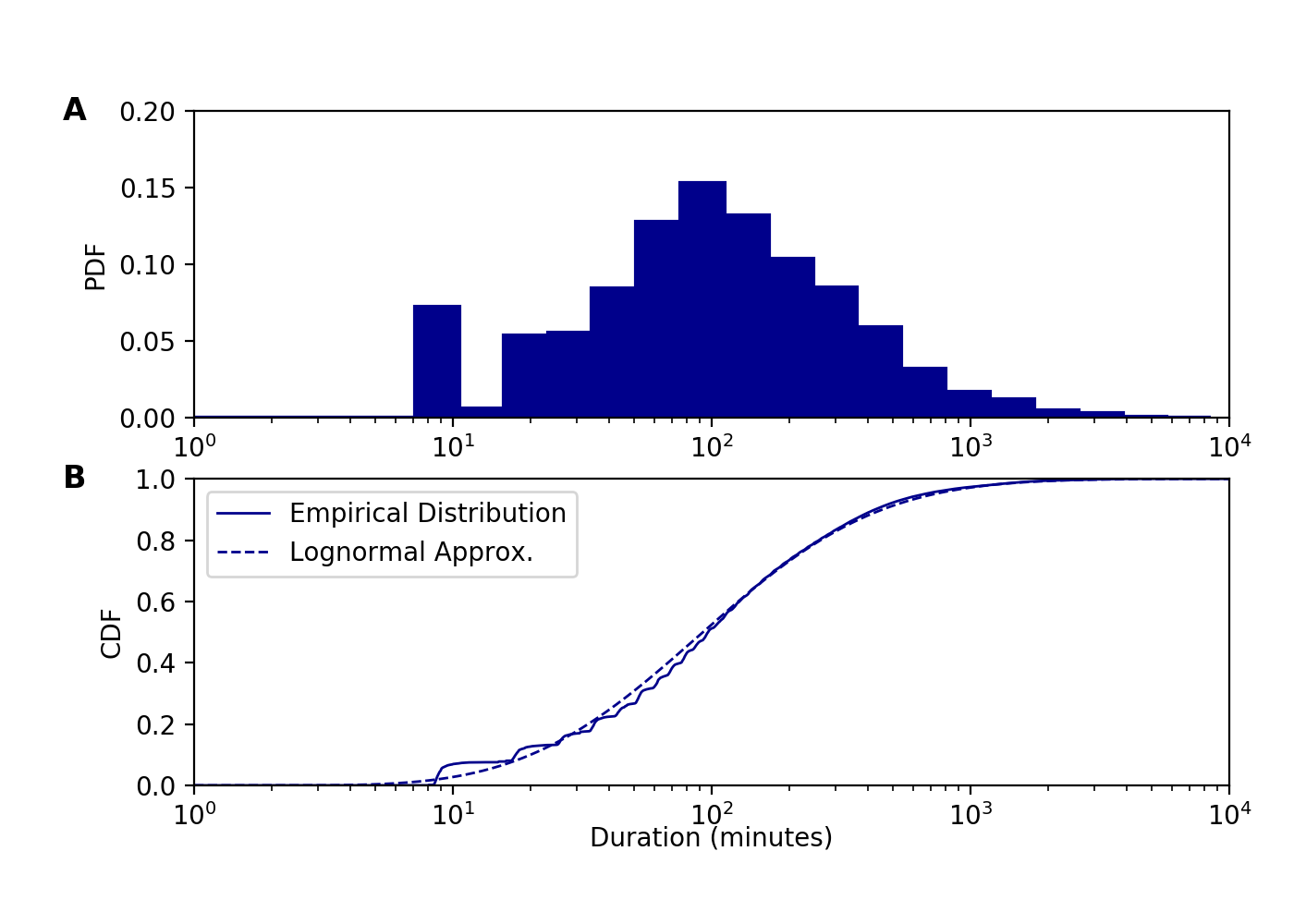}
\end{center}
\caption{(A) Probability and (B) cumulative distributions of power interruption duration (in minutes). 10\% of the observed interruptions last fewer than 10 minutes. The dashed line in (B) shows that the distribution is approximately log-normal with a mean of 95 minutes and a standard deviation of 3.}
\label{fig:duration}
\end{figure}

Figure \ref{fig:duration} shows the probability and cumulative distribution functions (or PDF and CDF), of power interruption duration.
We find that durations follow an approximately log-normal distribution with geometric mean and standard deviation equal to 95 and 3 minutes, respectively.
While we explored other parametric distributions, we find the log-normal distribution provides the best overall fit to the data.

Though short-duration power interruptions ($<$30 minutes) deviate from the log-normal distribution, this observation may simply be a construct of our dataset.
Discrete sampling intervals mean that the duration of power interruptions is not well characterized unless the duration is much longer than our sampling frequency.

\emph{Implication 1:} The distribution of power interruption duration can be well characterized by only two parameters: one describing the central value of the distribution and another describing the variability among observations.

\emph{Implication 2:} The range of typical restoration times spans several orders of magnitude.

\subsubsection{Discussion}
The heterogeneity we observe is not evidenced in public data reporting the mean duration (CAIDI), but not the variance.
Implication 1 suggests that the heterogeneity is well approximated by a parametric distribution.
Publicly reporting both the mean and the variance would increase transparency into the distribution of restoration times and the relative frequency of long-duration power interruptions.

Furthermore, because the distribution is not symmetric, the mean (CAIDI) is skewed by a relatively small number of power interruptions with durations that are orders of magnitude higher than average.
In our data, about 70\% of observations fall below the arithmetic mean and about 30\% fall above; these proportions are characteristic of a log-normal distribution.
The arithmetic mean is 180 minutes or nearly twice the geometric mean (which describes the center of the distribution).
The center of the distribution could be more accurately represented if the median or geometric mean were reported in addition to the arithmetic mean.
The geometric mean would better represent the typical duration, without being affected by long-duration interruptions at the tail of the distribution.

These new metrics could support efforts to more accurately compute the damages due to power interruptions, as survey data consistently reports that damages scale with interruption duration (for example, see \cite{sullivan2015}, \cite{chang2007} and \cite{carlsson2008}).
Works in the literature to date typically use CAIDI to represent interruption duration \cite{lacommare2006, larsen2016}.
Our results show that this approximation overestimates damages for the 70\% of power interruptions that are shorter than the mean, while underestimating damages (potentially by orders of magnitude) for the 30\% of power interruptions that are longer than the mean.
It is unclear whether more granular information would increase or decrease overall damage estimates such as those reported in \cite{larsen2016} and \cite{lacommare2006}.

Implication 2 underscores the need for further research characterizing damages due to long-duration power interruptions, as damage functions reported in Sullivan et al \cite{sullivan2015} (and applied in \cite{larsen2016,lacommare2006}) are only valid for power interruptions shorter than 8 hours.
Based on our records, nearly 8\% of power interruptions include some customers who remain without service for more than 8 hours.
These long-duration interruptions are potentially much more damaging than short interruptions.

Implication 2 also raises questions about whether certain customers are more likely to experience long-duration interruptions than others. 
The literature suggests that this may be the case, as policies targeting metrics like SAIDI and SAIFI can favor upgrades in high-density areas where service reliability is already high \cite{aguero2009}.
However, if studies could show that long-duration interruptions contribute a higher share to overall damages, policies targeting damage reductions instead of SAIDI and SAIFI would likely favor measures that benefit customers in underperforming regions of the grid. 

However, the lowest performing areas of the grid may be remote areas with few customers per line-mile.
The implication is that there may be high costs to improve service for a relatively small number of customers.
Where infrastructure upgrades are not cost effective, backup power systems or local generation may provide a cheaper alternative to achieve similar damage reductions. 
Today, it is customary for individual grid customers to shoulder the costs of backup power systems while ratepayers shoulder the costs of infrastructure upgrades. 
Our results raise questions about who should pay for backup power, and whether incorporating distributed generation into the planning process could lead to more efficient investment in grid performance, or more equity in how the burden of power interruptions is distributed among different grid customers.

\FloatBarrier
\subsection{Variability in interruption size}
\label{sec:results:size}
\subsubsection{Results}

\begin{figure}[ht]
\begin{center}
\includegraphics[width=0.8\textwidth]{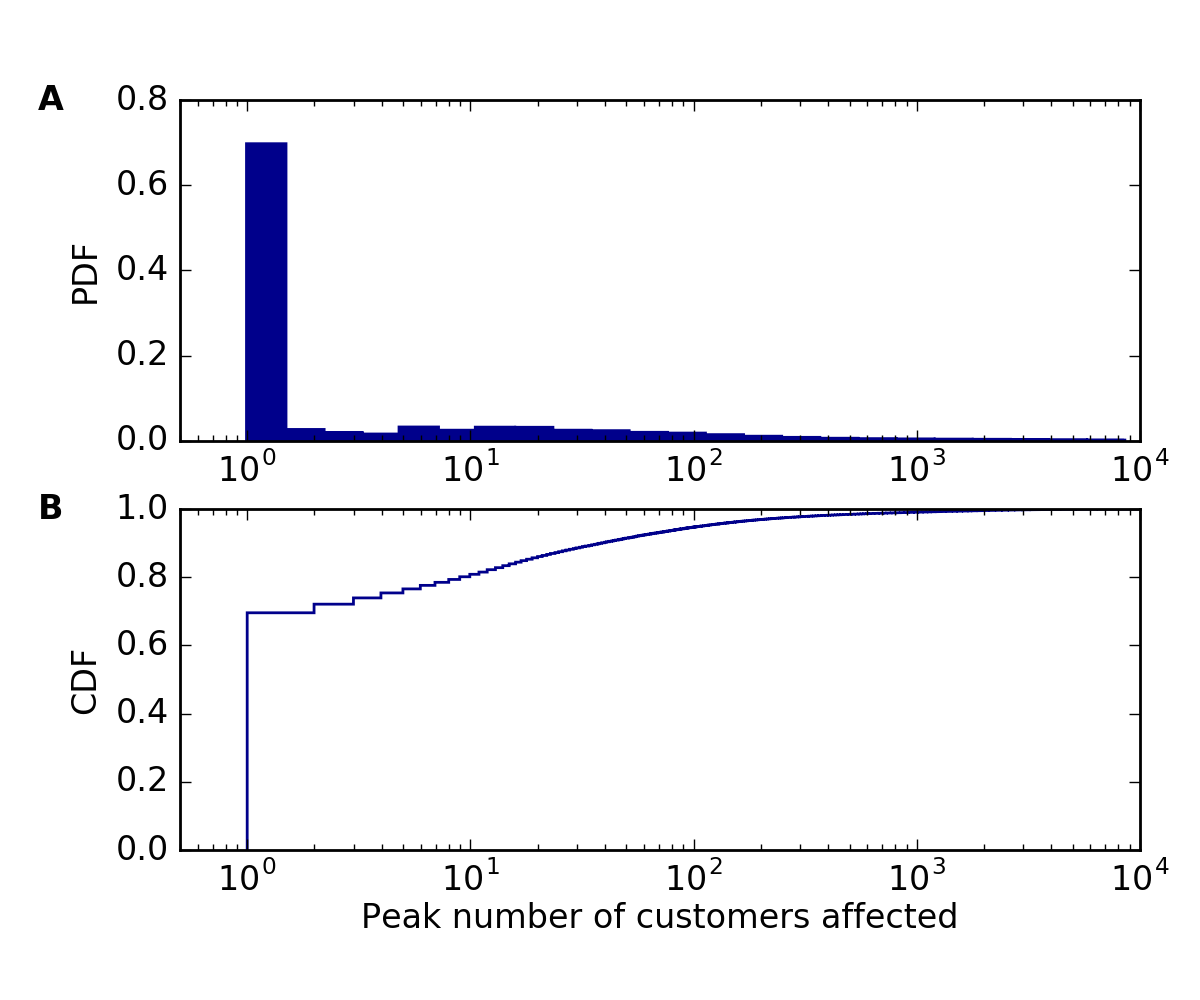}
\end{center}
\caption{(a) Probability and (b) cumulative distribution of the peak (or maximum) number of customers affected by individual power interruptions. 70\% of observed power interruptions affect at most one customer; no single interruption affected enough customers (50,000 or more) to be considered an emergency incident.}
\label{fig:peak_customers}
\end{figure}

Next we examine the PDF and CDF of the maximum (or peak) number of customers recorded during each individual power interruption (Figure \ref{fig:peak_customers}).
Of these power interruptions, 70\% affect only one customer and 80\% affect no more than 10 customers.
By our definition of a power interruption, we record no single interruption that affected enough customers (50,000 or more) to be classified as an emergency incident based on size alone.

However, DOE records include several emergency incidents that are spatially and temporally aligned with our dataset.
The reason we do not directly observe these large-scale events is that the definition of an emergency incident differs from our own definition of a power interruption.
An emergency incident can include power interruptions that occurred at multiple locations so long as they are linked by a common cause (for example a contingency event, or severe weather).

Our dataset does not include enough information to draw causal relationships between interruptions.
Instead, we look for possible emergency incidents by examining time intervals where at least 50,000 customers were without power in the service territory as a whole.
If these interruptions are causally linked, then we expect them to be included in the emergency incident data.

\begin{figure}[t]
\begin{center}
\includegraphics[width=0.8\textwidth]{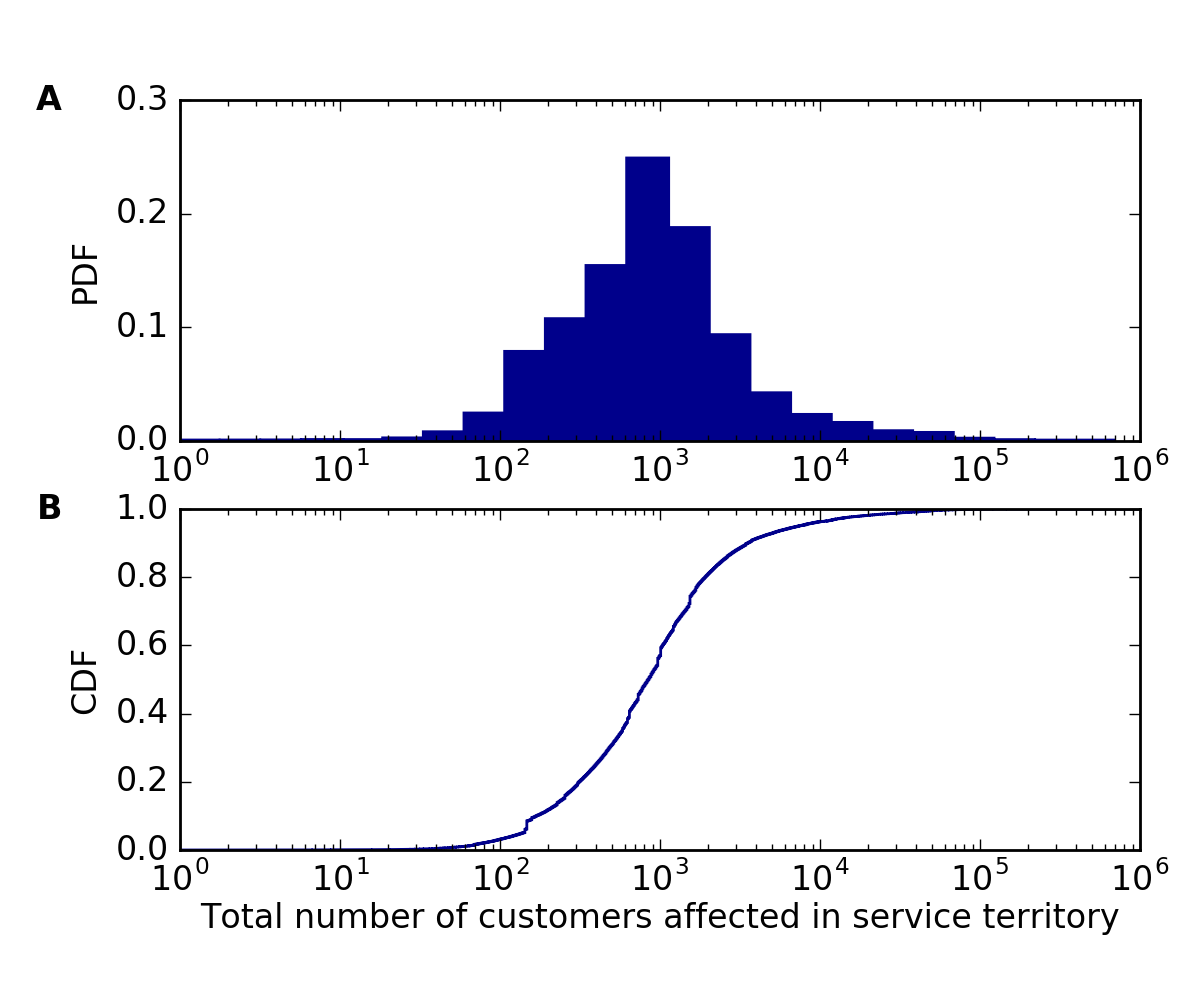}
\end{center}
\caption{(a) Probability and (b) cumulative distribution of the total number of customers without service in the entire service territory at each observation time. 600 observations (0.5\% of observations) report that the total number of customers without service exceeds 50,000 customers.}
\label{fig:total_customers}
\end{figure}

Figure \ref{fig:total_customers} shows the PDF and CDF of the total number of customers affected in the service territory across all of the observations we record.

We recorded 600 observations during which time more than 50,000 customers were without power for at least one hour.
These observations span 14 continuous time intervals, 11 of which coincide with emergency incidents reported in the DOE dataset.
Our dataset does not include enough information to determine what differentiates the remaining 3 incidents from the 11 that were subject to DOE reporting requirements.

Over all three years, these 14 time intervals collectively account for 0.5\% of the observations we record, 8\% of the power interruptions we observe, 40\% of SAIDI and 13\% of SAIFI.
Ten of these intervals occurred in 2017 (including all three that were not reported in the DOE data).
If we exclude observations recorded after January 1 2017, emergency incidents account for only 6\% of SAIFI (see Implications 2 and 3 below).

Table \ref{tab:emergency_incidents} lists the approximate time of each incident, the peak number of customers affected, and the number of customers reported in the DOE dataset.
We note that despite methodological differences in how we calculate the number of customers affected by each incident, we arrive at remarkably similar values as those reported the DOE dataset (see Implications 5 and 6 below).

\begin{table}[t]
\caption{Time and size of power interruptions affecting a peak number of 50,000 or more grid customers, and corresponding total number of affected customers, as reported to DOE. ``Unknown'' indicates that an incident was reported to DOE but that the number of customers affected was reported to be unknown; ``Not Reported'' indicates that an incident was not reported to DOE.}
\begin{centering}
\begin{tabular}{p{0.5in} p{0.5in} m{1.75in} m{1.75in} }
\textbf{Year} & \textbf{Season} & \textbf{Peak customers recorded} & \textbf{Total customers reported} \\ \hline
2014 & Winter & 120,000 & Unknown \\
2014 & Winter & 74,000 & 84,000 \\
2015 & Winter & 64,000 & 65,000 \\
2015 & Fall & 57,000 & 56,000 \\
2017 & Winter & 104,000 & 106,000 \\
2017 & Winter & 78,000 & 87,000 \\
2017 & Winter & 74,000 & 75,000 \\
2017 & Winter & 61,000 & 64,000 \\
2017 & Winter & 168,000 & 169,000 \\
2017 & Winter & 52,000 & Not Reported \\
2017 & Winter & 76,000 & Not Reported \\
2017 & Winter & 57,000 & Not Reported \\
2017 & Spring & 130,000 & 100,000 \\
2017 & Spring & 90,000 & 88,000 \\ \hline
\end{tabular}
\end{centering}
\label{tab:emergency_incidents}
\end{table}

\emph{Implication 3:} Year-to-year differences in the size and frequency of large-scale interruptions can have major implications on SAIFI.

\emph{Implication 4:} Emergency incident data provides only a limited sample of the power interruptions that occurred. 
Furthermore, the sample is not representative of overall grid performance, as it only includes the most severe interruptions.

\emph{Implication 5:} While we cannot with certainty confirm the accuracy of the data we collect, we calculate approximately the same number of customers affected as is reported in the emergency incident data.
This observation suggests that our data are indeed representative of the interruptions that occurred.

\emph{Implication 6:} Despite methodological differences in how we estimate the number of customers affected by each incident, we observe less than a 1\% overall difference in the number of customers we compute compared with those reported in the DOE dataset.
This result indicates that day-to-day power interruptions have a relatively small impact on the overall number of customers affected compared with large-scale disturbances.

\subsubsection{Discussion}
The bias we observe (Implication 2) raises questions about whether analyzing comprehensive power interruption data could reveal patterns that are characteristically different from the patterns Hines et al \cite{hines2009} observe in emergency incident data.
For example, large-scale power interruption data likely exhibit seasonal patterns more aligned with severe weather events (such as hurricanes) than weather patterns with more nuanced reliability implications.
In Section \ref{sec:results:patterns}, we explore the implications of bias by reporting on temporal patterns observed in our dataset and compare our findings with observations reported in Hines et al \cite{hines2009}.

The observation of bias also raises questions about how to group power interruptions based on criteria that provide more actionable insights than grouping them simply by size.
For example, brief descriptions of vulnerabilities that have led to emergency incidents data include severe weather and cascading failure.
However, the same causal factors can also result in power interruptions that affect fewer than 50,000 customers and may not be reported as emergency incidents.
Examining how frequently emergency incidents are attributed to a particular cause does not necessarily tell how severe a particular vulnerability really is, or how much the system will benefit from policies for mitigating it.

Novel methods for classifying power interruptions could yield new insights that emergency incident data do not provide.
For example, grouping power interruptions based on the environmental stressors that caused them (e.g., high wind speed or lightning) could inform decisions about which vulnerabilities are most urgent to address.
Further research is needed to determine the most meaningful ways to classify power interruptions, and to develop statistical methods for doing so.

\FloatBarrier
\subsection{Temporal patterns in SAIDI, SAIFI and CAIDI}
\label{sec:results:patterns}
\subsubsection{Results}
\begin{figure}[t]
\begin{center}
\includegraphics[width=0.8\textwidth]{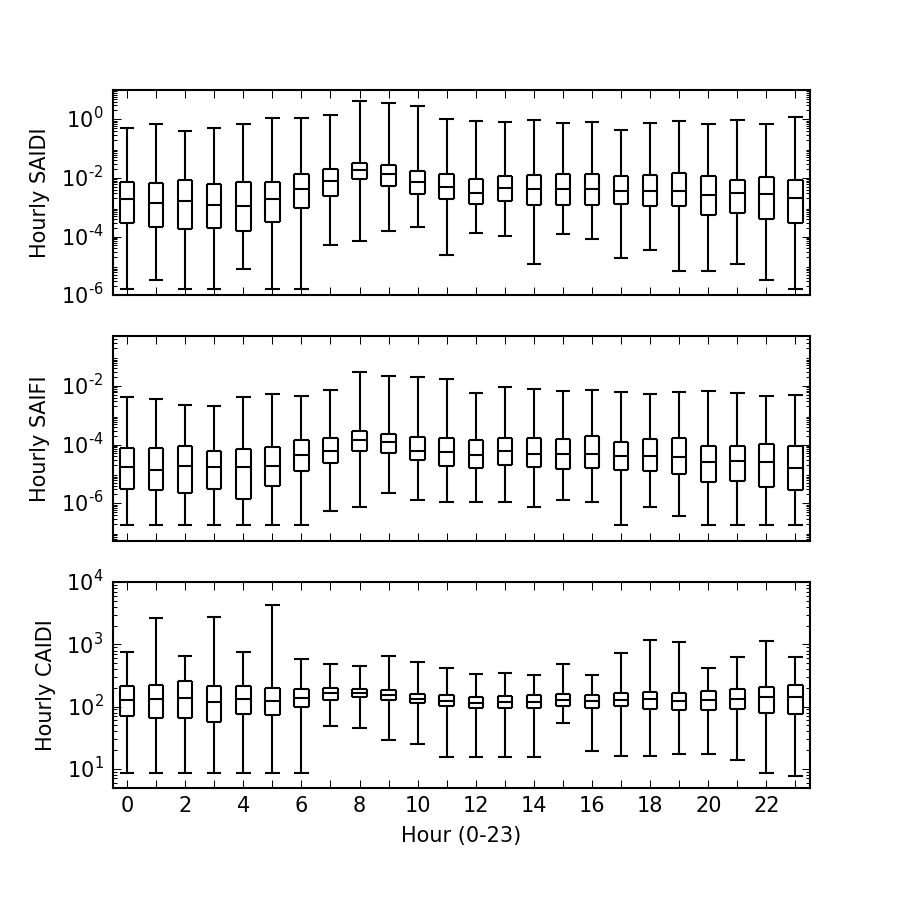}
\end{center}
\caption{Box-and-whisker plot showing the distribution of SAIDI, SAIFI and CAIDI computed in hourly increments and grouped by time of day. Whisker endpoints denote the 5th and 95th percentiles, box endpoints denote the 25th and 75th percentiles, and the horizontal line inside each box denotes the median value.}
\label{fig:diurnal_boxplots}
\end{figure}

Figure \ref{fig:diurnal_boxplots} shows box-and-whisker plots describing the distribution of SAIDI, SAIFI and CAIDI calculated in hourly increments and grouped by time of day.
The data show that both SAIDI and SAIFI typically increase in the morning, peak at about 08:00AM, and decrease over the remainder of the day.
The only appreciable change in CAIDI is that variability decreases among power interruptions that begin between 08:00 and 10:00AM.
One possible explanation for this trend is that planned maintenance typically occurs during those hours, a hypothesis which is supported by the observation that the number of power interruptions (SAIFI) is an order of magnitude higher in the morning than at night.

\begin{figure}[t]
\includegraphics[width=\textwidth]{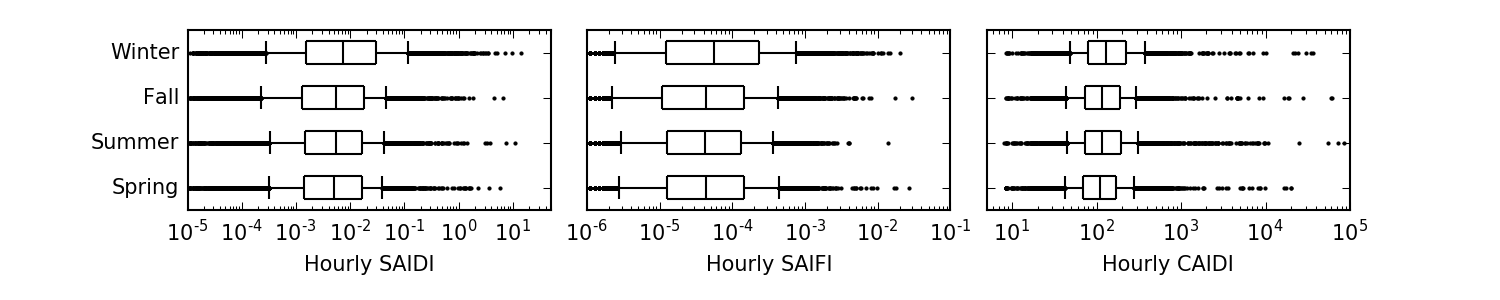}
\caption{Box-and-whisker plot showing the distribution of SAIDI, SAIFI and CAIDI computed in hourly increments and grouped by season. Whisker endpoints denote the 5th and 95th percentiles, box endpoints denote the 25th and 75th percentiles, and the vertical line inside the box denotes the median value, and points denote observations outside the 5th and 95th percentiles for each season.}
\label{fig:seasonal_hourly_boxplots}
\end{figure}

\begin{figure}[t]
\includegraphics[width=\textwidth]{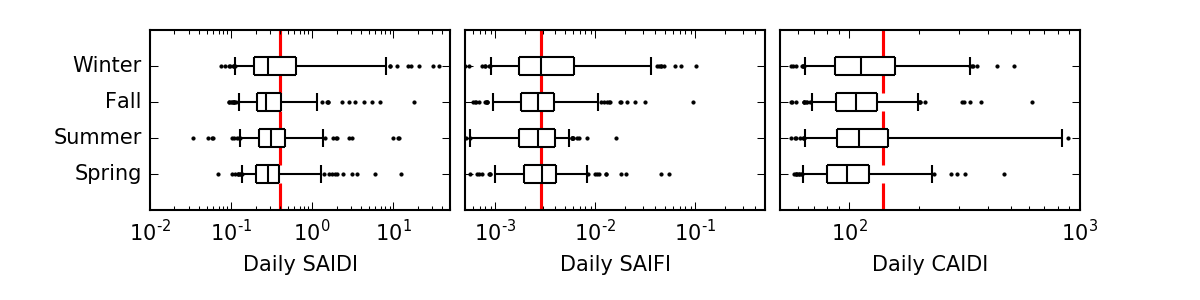}
\caption{Box-and-whisker plot showing the distribution of SAIDI, SAIFI and CAIDI computed in daily increments and grouped by season. Whisker endpoints denote the 5th and 95th percentiles, box endpoints denote the 25th and 75th percentiles, the vertical line inside each box denotes the median value, and points denote observations outside the 5th and 95th percentiles for each season. The red lines denote the average daily values of each metric reported to EIA in 2015.}
\label{fig:seasonal_daily_boxplots}
\end{figure}

Figures \ref{fig:seasonal_hourly_boxplots} and \ref{fig:seasonal_daily_boxplots} show box-and-whisker plots of SAIDI, SAIFI and CAIDI by season.
We calculate each reliability metric by grouping observations into hourly (Figure \ref{fig:seasonal_hourly_boxplots}) and daily increments (Figure \ref{fig:seasonal_daily_boxplots}).

Figure \ref{fig:seasonal_hourly_boxplots} shows no evidence of seasonal heterogeneity in hourly metrics.
However, Figure \ref{fig:seasonal_daily_boxplots} shows that the extreme values of daily SAIDI and SAIFI are one to two orders of magnitude higher in winter than in other seasons.
The 5 days with the most severe values of SAIDI account for 20\% of overall SAIDI; the most severe 30 days account for 50\% of overall SAIDI (see Implication 8).

\emph{Implication 7:} Power interruptions are not homogeneously distributed over the course of the day.
Rather, patterns in SAIDI suggest that interruptions are more likely to occur at certain times of day. 
We see no clear evidence to suggest that interruption duration (CAIDI) follows a similar characteristic diurnal pattern.

\emph{Implication 8:} In our dataset, seasonal and annual changes in reliability metrics are largely attributable to a small number of days with particularly severe power interruptions.

\emph{Implication 9:} Daily reliability metrics exhibit seasonal heterogeneity that hourly metrics do not.
This suggests seasonality is driven by interruptions that intensify over the course of several hours, rather than interruptions that occur instantaneously.

\subsubsection{Discussion}
The seasonal and diurnal patterns Hines et al \cite{hines2009} observe in emergency incident data differ from the patterns we report.
For example, we record the most power interruptions during winter months, while Hines et al report that the frequency of emergency incidents increases during both summer and winter months.
We also observe that most power interruptions begin in the morning, while Hines et al report that most emergency incidents began in the afternoon.

One explanation for these differences is that our dataset focuses on a single utility service territory, while Hines et al report on emergency incidents observed nationwide.
Regional differences in weather patterns could lead to characteristically different seasonal patterns in grid performance. 
For example, the utility we examine (which is located in the Western United States) may not be exposed to particularly severe weather during the summer.
However, utilities in other parts of the country may be vulnerable to lighting storms, tornadoes or hurricanes that occur during the summer.

Another explanation for differences between the two datasets is that the causal factors that lead to emergency incidents follow characteristically different temporal patterns than the factors that lead to small-scale power interruptions.
For example, say we could confirm that the diurnal pattern we observe is attributable to scheduled maintenance.
Because scheduled maintenance likely does not lead to emergency incidents, we would not expect emergency incident data to exhibit the same diurnal patterns as we observe in Figure \ref{fig:diurnal_boxplots}.

Characterizing temporal patterns can provide insights relevant to (1) grid operations/planning, (2) damage estimation, (3) risk assessment, and (4) identifying the drivers of year-to-year differences in grid performance.

First, Hines et al discuss how patterns observed in large-scale power interruptions could inform policy decisions for improving grid performance.
Yet responding efficiently and effectively to large-scale power interruptions is only one component of maintaining a reliable and resilient grid. 
Low-cost measures may also reduce the duration or frequency of small-scale power interruptions which cumulatively contribute approximately the same amount to annual SAIDI as large-scale power interruptions.
Though public datasets do not contain sufficient information to be able to characterize these patterns, doing so could help in identifying opportunities for improving performance.

Second, results in the literature show that damages are related to the time of day and time of year a power interruption occurs \cite{carlsson2008, chang2007, sullivan2015}.
For lack of more granular information, both LaCommare and Eto \cite{lacommare2006} and Larsen \cite{larsen2016} assume that power interruptions are homogeneously distributed with respect to time.
Replicating these studies with data describing temporal heterogeneity in SAIDI, SAIFI and CAIDI would lead to more accurate damage estimates than those currently reported.

Third, there is discussion in the literature regarding the growing penetration of electric heating, cooling, and transportation loads, and how these loads will affect electric power systems.
One under-explored research area is how growth of high-value loads could change the severity of damages, particularly if patterns in SAIDI and SAIFI coincide with times when these loads are most needed.
Case studies of power interruptions in Canada and Sweden (where penetration of electric heating is high) reported loss of electric heating loads to be a leading cause of damages primarily because both interruptions occurred during severe ice storms \cite{carlsson2008,chang2007}.
Research examining the interplay between grid performance and end use electrification could help in anticipating emerging power system vulnerabilities before high-damage power interruptions occur.

Finally, a deeper understanding of seasonal trends could also advance efforts to understand how and why grid performance is changing over time.
Studies reporting that grid performance has been decreasing \cite{larsen2015, amin2008, hines2009} remain inconclusive as to what is driving this trend.
For example, year-to-year changes in grid reliability could be due to weather patterns, or to aging grid infrastructure.
Caswell et al \cite{caswell2011} discuss methods for normalizing reliability metrics to control for weather in order to isolate changes due to other factors.
They point out that to do so, the spatio-temporal granularity of the reliability metrics calculated must be similar to the granularity of the weather phonemona that caused the change in grid performance.
Since public metrics are typically reported annually and over large service territory, they are not granular enough to support weather normalization.
However, controlling for weather-related differences is a necessary step towards evaluating how much of the variability in performance is due to weather compared with other factors such as operational, maintenance, or investment decisions.
Discussion of these methods in the literature to date is largely conceptual, and quantitative research is needed to explore how much aggregation is appropriate.

% \FloatBarrier
\subsection{Geographic variability in interruption duration}
\label{sec:results:geography}

\subsubsection{Results}
Finally, we examine geographic variability in CAIDI (or the average restoration time) across different regions of the service territory.
We map the reported point locations to ZIP codes and compute CAIDI for interruptions observed in each ZIP code.
We note that calculating SAIDI and SAIFI at this geographic resolution is not possible using the current dataset, as $C_{tot}$ is known only for the service territory as a whole.
However, $C_{tot}$ is not needed to compute CAIDI, as CAIDI is a ratio between SAIDI and SAIFI (where both quantities are normalized by the same value $C_{tot}$).

\begin{figure}[t]
\includegraphics[width=\textwidth]{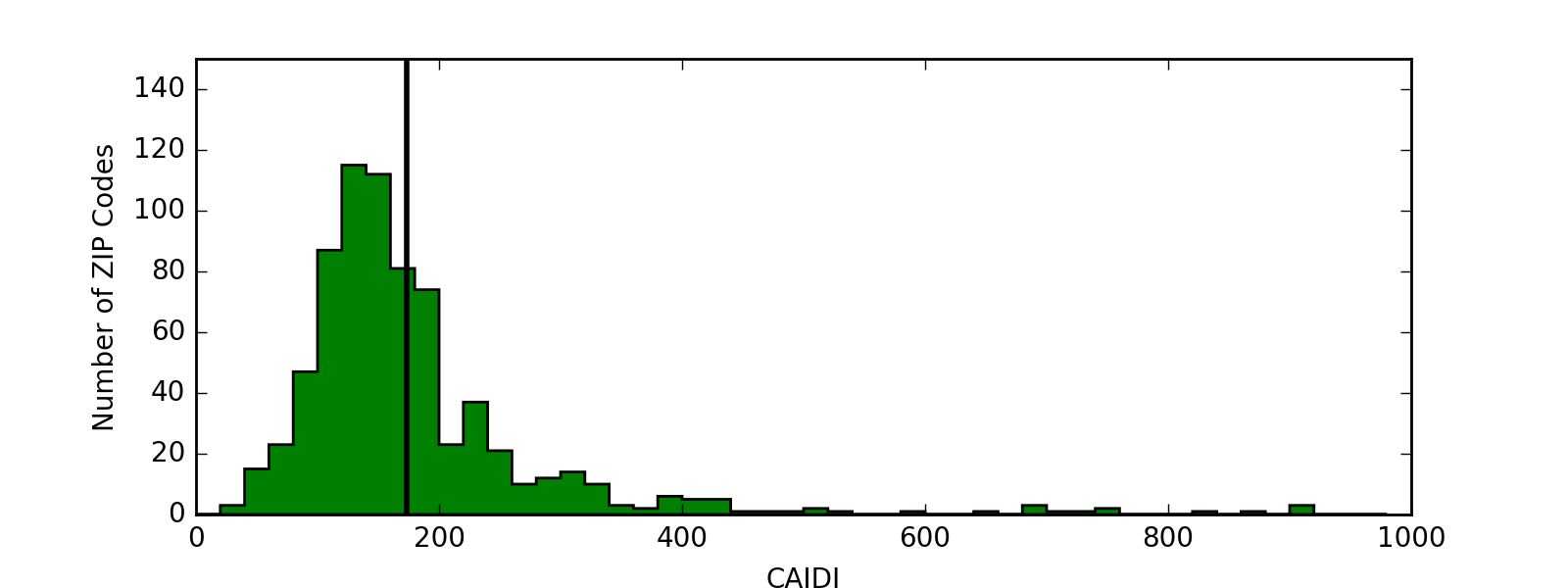}
\caption{Distribution of annual CAIDI for each ZIP code in the service territory. The vertical line denotes overall system CAIDI (170 minutes).}
\label{fig:caidi}
\end{figure}

Figure \ref{fig:caidi} shows the distribution of CAIDI for 700 ZIP codes in the service territory, excluding ZIP codes with fewer than 50 power interruptions.
Across all 700 ZIP codes, we calculate that CAIDI is 170 minutes (denoted by the vertical line).
Within individual ZIP codes, CAIDI ranges from 10 to 1,000 minutes.

\emph{Implication 10:} Customers in different parts of the service territory experience characteristically different restoration times.

\emph{Implication 11:} Because they are averaged over large geographic regions, system-wide metrics such as CAIDI (and perhaps SAIDI and SAIFI as well) miss a great deal of variability in the typical characteristics of power interruptions that customers in different subsets of the service territory experience.

\subsubsection{Discussion}

Characterizing geographic heterogeneity in CAIDI could lead to potentially large changes in damage estimates reported in the literature.
Damage functions reported in Sullivan et al \cite{sullivan2015} are sensitive to the duration of a power interruption, as well as the mix of customers affected.
For lack of more granular data, works in the literature to date assume that SAIFI and CAIDI are homogeneous throughout the service territory.
In other words, these studies assumed that in a given year, all customer types experience the same number and duration of power interruptions.
If it were shown (for example) that industrial customers experience characteristically shorter restoration times than residential customers, damage estimates would decrease because commercial and industrial customers make up the largest share of these damages \cite{lacommare2006}.

Reporting reliability metrics for different subsets of customers (e.g., customer type, ZIP code, rate class, or service level) would also unlock opportunities to explore a wide range of policy questions that to our knowledge remain unanswered.
For example, reliability metrics reported by customer type would allow us to more accurately assess damages.
Furthermore, reliability metrics reported by ZIP code could provide insight into whether the costs of power interruptions are being equitably distributed among customers, or whether policy intervention is needed to redistribute these costs.
Detailed consideration of these (and other) use cases would be needed to determine how much granularity is warranted in publicly reported data, or whether granular data should be made available to researchers on a case-by-case basis.

Characterizing heterogeneity in power interruption characteristics could also inform data-driven methods for optimizing investment in performance upgrades.
For example, Larsen \cite{larsen2016} describes a cost-benefit framework for evaluating if and where undergrounding distribution infrastructure is cost effective.
The author finds that undergrounding is typically (though not always) cost effective in urban but not in rural areas.
However, that result is sensitive to the assumption that restoration times are homogeneous throughout the service territory,  and that they are the same for all customers.

To illustrate how these assumptions could affect damage estimates, consider hypothetical results showing (1) that grid customers in urban areas experience fewer and shorter power interruptions than customers in rural areas, and (2) that there are more commercial and industrial customers located in urban areas than in rural areas.
After updating damage functions to account for heterogeneity in interruption duration (result 1), the benefits of undergrounding urban power lines would likely decrease because urban customers typically experience restoration times that are shorter than CAIDI.
On the other hand, updating damage functions to account for heterogeneity in the customer mix (result 2) would increase the benefits of undergrounding in certain areas, as commercial and industrial customers incur much higher damages than residential customers \cite{sullivan2015}.
Understanding how the characteristics of power interruptions differ for different groups of customers is critical to informing undergrounding policies.

This example illustrates how spatially resolved power interruption data can help in developing more accurate models for evaluating the costs and benefits of a particular policy for investing in grid performance.
Such policies can support decisions to invest in system upgrades if and where higher levels of service reliability are warranted.

\FloatBarrier
\section{Conclusion and Policy Implications}
\label{sec:conclusion}

The main contribution of this work is to show that power interruption characteristics exhibit substantial geographic and temporal heterogeneity that are not evidenced in publicly available datasets.

The heterogeneity we observe sheds light on a number of important policy questions.
Detailed analysis exploring these questions will require high-resolution data, as aggregate metrics provide limited insight into the characteristics of the power interruptions that individual grid customers experience.
We discuss a wide range of analysis questions that could leverage high-resolution data to advance how we evaluate, regulate, and invest in grid performance.

We contextualize our results by examining how granular data could change results reported in the literature to date, and show that limited access to high-resolution data has been a barrier to research progress.

As energy sources, grid infrastructure, and electricity end uses evolve, the causes and implications of power interruptions are also likely to change.
To ensure that power systems are able to support electrification of new end uses such as transportation or heating without adverse consequences, granular research is needed to examine the drivers of regional and year-to-year differences in performance.
Research is also needed to assess adequacy and equity of grid performance today, and to confirm that both the burden of underperformance and the costs of improving performance are equitably distributed among ratepayers.
Examining these questions could help us to better understand and quantify grid performance, and ensure that policymakers are leveraging state-of-the-art data and methods to make the most informed decisions available to them.

Although high-resolution data suitable for exploring these questions do exist, these data are proprietary and are not readily made available to researchers conducting policy analysis.
However, the consequences associated with leaving unanswered questions related to current and future grid performance could be severe.
To ensure that grid performance is adequate to support the needs of a society that relies heavily on electricity, it is our recommendation that high-resolution power interruption data be made more readily available to support data-driven reliability research.

\section*{Acknowledgements}
\label{sec:acknowledgements}

The work described in this paper was funded by the U.S. Department of Energy, Office of Electricity Delivery and Energy Reliability in accordance with the terms of Lawrence Berkeley National Laboratory Contract No. DE-AC02-05CH11231. Our funders had no involvement in the preparation or submission of this manuscript. We would like to acknowledge Nathan Addy for developing the software infrastructure for collecting the data that supported this work.

\section*{References}
\bibliographystyle{ieeetr}
\bibliography{refs}

\end{document}